\documentclass{article}
\usepackage[utf8]{inputenc}
\usepackage{amsmath,amssymb,bbm}
\usepackage{url}
\usepackage{preamble}

\title{Necessary Conditions in\\Multi-Server Differential Privacy}
\author{Albert Cheu \& Chao Yan\thanks{Email ac2305@georgetown.edu and cy399@georgetown.edu}\\Department of Computer Science\\Georgetown University}

\begin{document}

\maketitle

\begin{abstract}
We consider protocols where users communicate with multiple servers to perform a computation on the users' data. An adversary exerts semi-honest control over many of the parties but its view is differentially private with respect to honest users. Prior work described protocols that required multiple rounds of interaction or offered privacy against a computationally bounded adversary. Our work presents limitations of non-interactive protocols that offer privacy against unbounded adversaries. We show these protocols demand exponentially more samples for some learning and estimation tasks than centrally private counterparts. This means performing as well as the central model requires interactivity or computational differential privacy, or both.
\end{abstract}

\section{Introduction}
Following the seminal work by Dwork, McSherry, Nissim, and Smith \cite{DMNS06}, much research in differential privacy takes place in the central model. This assumes owners of data are willing to give their data to a central analysis server---an \emph{analyst} for short---who runs a differentially private algorithm and reports the output to the (adversarial) world. Such an algorithm will guarantee that, loosely speaking, its output will not leak much information about any individual who gave their data. But the analyst could run other algorithms, so leaks and data misuse can still occur.

To keep data out of an analyst's hands, prior research has produced a variety of alternative models. A well-studied example is the local model. Here, \emph{users} of mobile phones or web browsers run differentially private algorithms on their data and send the resulting messages to the analyst \cite{Warner65, KLNRS08}. The local nature of the randomization ensures privacy for any user even when the analyst and all others users are corrupted by the adversary. An inherent limitation of such a protocol is that the privacy noise from the users significantly weakens the signals they are meant to send; Kasiviswanathan, Lee, Nissim, Raskhodnikova, and Smith showed that locally private parity learning demands exponentially more samples than centrally private parity learning \cite{KLNRS08}.

A line of work augments local protocols with a shuffler, an intermediary that applies a random permutation on user messages before sending the result to the analyst \cite{BittauEMMRLRKTS17, CSU+19}. The anonymity offered by the shuffler acts as a second layer of protection, atop the local randomization. The shuffler can be securely instantiated via anonymous broadcast protocols (e.g. Eskandarian \& Boneh \cite{EB21}) or classic mixnets (see Chaum \cite{Chaum81}).

We focus on an alternative relaxation of the local model that appears in various forms in prior work \cite{Steinke20, AppleGoogle21, Talwar22, BGGKMRS22}: instead of just one analysis server, we assume there are $k\geq 2$ servers who share responsibility in processing user messages. An adversary can corrupt all but one of the users (as in the local model) and a large fraction of the servers, identities unknown.\footnote{Steinke \cite{Steinke20} and Talwar \cite{Talwar22} describe protocols that ensure privacy holds when $k-1$ servers are corrupt. Our results apply to protocols with a weaker guarantee.} The adversary observes the messages received by parties it corrupts; we would like this view to change only slightly when an honest user changes their contribution. We remark that this multi-server model subsumes the shuffle model, since Eskandarian \& Boneh create a multi-server protocol that performs anonymous broadcast \cite{EB21}.

That construction relies on the assumption that the adversary is computationally bounded. As noted by Steinke \cite{Steinke20}, such an assumption also implies accurate simulation of centrally private algorithms via secure multiparty computation. Meanwhile, classic work in central and local privacy allow \emph{unbounded adversaries}. The natural line of thought is to explore the power of multi-server protocols when playing against that stronger class of adversary.

\begin{figure}
    \centering
    \includegraphics[width=0.6\textwidth]{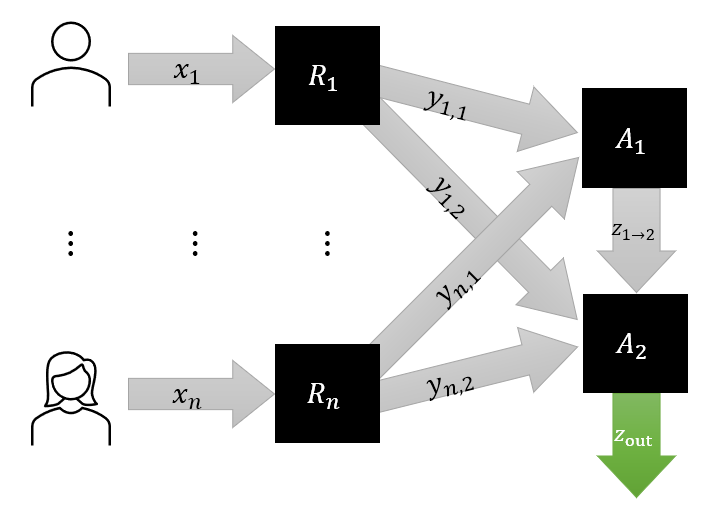}
    \caption{
    Diagram of a non-interactive two-server protocol. Users input their data into local randomizers, which send one message to each server. The first server produces a intermediary value, which the second server uses to produce the output. No party responds to messages (there are no cycles in the graph).
    }
    \label{fig:two-server}
\end{figure}

We are also interested in practical protocols: they should be computationally lightweight, consume low bandwidth, and take place over few rounds of interaction. We focus on interactivity. Ideally, our protocols would follow the \emph{non-interactive} communication pattern depicted in Figure \ref{fig:two-server}: each party produces one batch of outputs and never receives feedback. This is desirable because users' devices are not always connected and it is costly to keep servers online.

In Appendix \ref{apdx:example}, we sketch a non-interactive multi-server protocol that ensures differential privacy against an unbounded adversary. It accurately estimates the sum of bits held by users. But aside from this basic task, what can multi-server protocols compute if they must be non-interactive and ensure privacy against an unbounded adversary?

\subsection{Our Results \& Techniques}
\label{sec:results}

We show that the protocols of interest cannot perform some learning and estimation tasks without \emph{exponentially} more samples than centrally private algorithms. An example is feature selection. Section \ref{sec:lower-bounds} contains a formal theorem statement, but we give an informal version here:

\begin{thm}[Informal]
Consider the family of non-interactive multi-server protocols that offer $\eps$-differential privacy against unbounded adversaries that control $\leq \lceil k/2\rceil$ servers. If each user is drawn i.i.d. from a population and has $d$ binary features, any member of the protocol family requires $\Omega(\sqrt{d}/\eps)$ samples to select the most common feature. This is in contrast with the $O(\log (d)/\eps)$ sample complexity under central privacy, via the exponential mechanism.
\end{thm}


We also give lower bounds for parity learning and simple hypothesis testing, which likewise exhibit exponential gaps in sample complexity. Finally, we present a lower bound for uniformity testing, which is polynomially larger than the upper bound in the central model.

Our lower bounds have the following implication:
\begin{center}
To solve some learning and estimation tasks with as few samples as the central model, multi-server protocols must be interactive or only ensure computational differential privacy, or both.
\end{center}
An intriguing open question is whether interactivity alone suffices to learn parity with few samples. Our work also leaves open the possibility of sample-efficient, non-interactive parity learning with computational differential privacy.

\medskip

To arrive at our lower bounds, we take two high-level steps. First, we transform a multi-server protocol $\Pi$ into an \emph{internally private online algorithm} $A_\Pi$. Such an algorithm reads its input in a single for-loop and ensures that its internal state at the end of any single iteration respects differential privacy. This ensures that an adversary does not gain much advantage by intruding into the algorithm's memory. In our second step, we invoke lower bounds for internally private algorithms. These are implied by the work of Cheu and Ullman \cite{CU20} and Amin, Joseph, and Mao \cite{AJM19}.\footnote{Their lower bounds are phrased in terms of \emph{pan-private} online algorithms, which is technically a stricter constraint than internal privacy. But as noted in the thesis by Cheu \cite{Cheu21}, the lower bound arguments are also valid for internally private ones.}

Section \ref{sec:two-server-transform} describes the complete transformation for the two-server case; we briefly sketch the main ideas here. $A_\Pi$ processes its input stream $x_1,\dots,x_m$ in two batches. The first batch, labeled $[n]=\{1,\dots,n\}$, serves as the input to an execution of $\Pi$ (on $n$ samples): $A_\Pi$ simulates the construction of messages from every user $i\in[n]$ to server 1. It is tempting to also simulate the messages to server 2, but the pair of messages may be non-private.\footnote{Consider additive secret sharing amongst two servers: one share reveals nothing but the joint distribution is wholly dependent on the data value.} This motivates the second batch of samples, labeled $\{n+1,\dots , m \}$.

Naively, we could use the second batch to sample from the marginal distributions of the messages to server 2. But the message a user sends to server 2 could depend on the one it sent to server 1. So we instead rely on a technique by Joseph, Mao, Neel, and Roth \cite{JMNR19}: for each user $i$, $A_\Pi$ obtains a fresh sample from $\bD$ \emph{conditioned on having seen} the message from $i$ to server 1. This proxy for $i$'s data is then used to produce the message from $i$ to server 2. A technical hurdle arises from the fact that the desired conditional distribution requires the specification of $\bD$, which is unknown to the algorithm. But $A_\Pi$ approximates a sample from the conditional distribution by performing rejection sampling on the second batch of samples from $\bD$.

Joseph et al. developed their technique in the context of local protocols, where the adversary can see all of the messages that an honest user generates \cite{JMNR19}. In our multi-server model, the adversary can only see a subset of them. As a consequence, the simulator $A_\Pi$ will purge old messages from memory when they are no longer needed: once $A_\Pi$ simulates the message from $i$ to server 2, it erases the message from $i$ to server 1.

\subsection{Related Work}
\label{sec:related}

The whitepaper by Apple \& Google  presents an interactive protocol for exposure notification analytics \cite{AppleGoogle21}. It offers protection against bounded adversaries. Bell, Gascon, Ghazi, Kumar, Manurangsi, Raykova, and Schoppmann also describe an interactive protocol offering computational DP \cite{BGGKMRS22}. It provably achieves asymptotically optimal $\ell_\infty$ error for histogram estimation. Talwar \cite{Talwar22} and Steinke \cite{Steinke20} both describe information-theoretically secure protocols, though the vector summation protocol in \cite{Talwar22} assumes shared coins. Computational guarantees are offered by some protocols in \cite{Steinke20}.

See Table \ref{tab:comparisons} for a summary of the above works. To our knowledge, we are the first to prove lower bounds in our restricted version of the multi-server model.

\begin{table}
    \centering
    \begin{tabular}{|c|c|c|c|c|c|} \hline
         & No. & No. Corrupt & Bounded & Interactive? & Notes\\
         & Servers & Servers & Adversary? &  & \\ \hline
         \multirow{2}{*}{\cite{AppleGoogle21}} & \multirow{2}{*}{3} & \multirow{2}{*}{$\leq 2$} &  & & A protocol for histograms.\\
         & & & \multirow{2}{*}{Yes} &  & Users prove inputs are valid. \\ \cline{1-3}\cline{6-6}
         \multirow{2}{*}{\cite{BGGKMRS22}} & \multirow{2}{*}{2} & \multirow{2}{*}{$\leq 1$} &  & & A protocol for histograms\\
         & & &  &  & with optimal $\ell_\infty$ error \\ \cline{1-4}\cline{6-6}
         \multirow{2}{*}{\cite{Steinke20}} & & \multirow{4}{*}{$\leq k-1$} & Yes, & Yes & Protocols for counting, heavy- \\
         & & & for some &  & hitters, feature selection  \\ \cline{1-1}\cline{4-4} \cline{6-6}
         \multirow{2}{*}{\cite{Talwar22}} & & & Not if servers &  & A protocol for vector sum.\\
         & $k$ & & agree on coins &  & Rejects a user's vector if too big. \\\cline{1-1}\cline{3-6} 
         \textbf{This} &  & \multirow{2}{*}{$\leq \lceil k/2\rceil$}  & \multirow{2}{*}{No} & \multirow{2}{*}{No} & Lower bounds for feat. selection,\\
         \textbf{Work} & & &  &  & parity learning, and more\\ \hline
    \end{tabular}
    \caption{A non-exhaustive sampling of work on multi-server differential privacy. Most existing protocols are interactive and assume a bounded adversary. We are the first to give lower bounds.}
    \label{tab:comparisons}
\end{table}

McGregor, Mironov, Pitassi, Reingold, Talwar, and Vadhan \cite{MMPRTV10} present lower bounds in the \emph{two-party} model, as do Haitner, Mazor, Silbak, Tsfadia \cite{HMST22}. In that model, each party has direct access to half of all user data. The two parties interact over multiple rounds and each needs to ensure privacy against the other. Borrowing the visual language of Figure \ref{fig:two-server}, each user has only one arrow to one $A_j$ while a bidirectional arrow connects $A_1,A_2$. For the statistical estimation and testing problems we consider, the sample complexity in the central and two-party models are asymptotically identical. This is because an honest party can simply run a centrally private algorithm on their half of the samples.

Finally, we remark that our notion of privacy for online algorithms differs from the one found in the continual release literature (see Jain, Raskhodnikova, Sivakumar, and Smith \cite{JRSS21} \& citations within). There, the online algorithm outputs a value after every read. The stream of outputs must simultaneously respect differential privacy and serve as a good estimate of some function applied to each prefix (e.g. sum). In contrast, the online algorithms we construct only produce output at the end of the stream. Thus, our algorithms do not imply continual release algorithms. Moreover, we define privacy with respect to an arbitrary internal state chosen by the adversary, but a private continual release algorithm could conceivably maintain a non-private internal state.

\section{Preliminaries}

For any (possibly randomized) algorithm $M$ and distribution $\bD$ over inputs of $M$, $M(\bD)$ is shorthand for the distribution of $M(x)$ when $x\sim \bD$. For any pair of distributions $\bP,\bQ$, the expression $\sd{\bP}{\bQ}$ denotes the statistical (total variation) distance between the two.

We write $\bP \approx_{\eps,\delta} \bQ$ if, for all events $Y$, both of the following are true:
\begin{gather*}
\pr{}{\bP\in Y} \leq e^\eps\cdot \pr{}{\bQ\in Y} + \delta\\
\pr{}{\bQ\in Y} \leq e^\eps\cdot \pr{}{\bP\in Y} + \delta
\end{gather*}
In the case where $\delta=0$, we simply write $\bP\approx_\eps \bQ$.

\begin{fact}[Post-Processing]
\label{fact:post}
If $\bP \approx_{\eps,\delta} \bQ$, then for any algorithm $M$, $M(\bP) \approx_{\eps,\delta} M(\bQ)$
\end{fact}

Throughout this work, \emph{user} refers to a party who holds a single input value and \emph{server} refers to a party who does not hold any input.

\subsection{Non-interactive Multi-Server Protocols}

Here, we take the number of users to be $n$ and the number of servers to be $k>1$. A protocol $\Pi$ in the non-interactive multi-server model is specified by a tuple $(\{R_i\}_{i\in[n]}, \{A_j\}_{j\in[k]}, G)$. Each $R_i$ is a local randomizer run by user $i$ on their data; if all randomizers are identical, we simply write $R$. Meanwhile, $A_j$ is an algorithm run by server $j\in[k]$. Finally, $G$ is a $k$-node directed acyclic graph that determines the communication pattern between servers. We assume nodes (servers) are named according to a topological ordering. For brevity, we will drop the term ``non-interactive'' when discussing these protocols. Algorithm \ref{alg:multiserver} describes how $\Pi$ is executed.

\begin{algorithm}
\caption{The execution of a multi-server protocol $\Pi=(\{R_i\}_{i\in[n]}, \{A_j\}_{j\in[k]}, G)$ on input $\vec{x}$.}
\label{alg:multiserver}

    \For{$i\in [n]$}{
        User $i$ samples $(y_{i,1},\dots,y_{i,k})$ from $R_i(x_i)$
        
        User $i$ sends $y_{i,j}$ to server $j$, for all $j\in[k]$
    }
    
    \For{$j\in[k]$}{
        Server $j$ receives $y_{1,j},\dots,y_{n,j}$ from users and $z_{j'\to j}$ from servers $j'$ where $(j',j)\in G$
        
        
        \If{$j=k$}{
            The protocol's output is $z_\out \gets A_j(\{y_{i,j}\}_{i\in [n]}, \{z_{j'\to j}\}_{(j',j)\in G})$
        }
        \Else{
            Server $j$ samples $\{z_{j\to j''}\}_{(j,j'')\in G}$ from $A_j(\{y_{i,j}\}_{i\in [n]}, \{z_{j'\to j}\}_{(j',j)\in G})$
            
            Server $j$ sends $z_{j\to j''}$ to server $j''$
        }
    }

\end{algorithm}

An attack $\Phi$ is specified by a tuple $(C_u, C_s)$. $C_u\subset [n]$ is the set of corrupted users while $C_s\subset [k]$ is the set of corrupted servers. For any multi-server protocol $\Pi$, input $\vec{x}$, and attack $\Phi$, an adversary's \emph{view} in execution $\Pi_\Phi(\vec{x})$ is the random variable
$$
\view^{\Pi}_\Phi(\vec{x}) := (z_\out, \{z_{j\to j'}\}_{(j,j') \notin \overline{C_s} \times \overline{C_s}}, \{y_{i,j}\}_{(i,j)\notin \overline{C_u} \times \overline{C_s}}, \{x_i\}_{i\in C_u})
$$
That is, an adversary attacking $\Pi$ with $\Phi$ can observe the output of the protocol, all messages except those between honest parties, and the data of corrupted users.

For differential privacy to be satisfied, the adversary's view must be insensitive to any one user.
\begin{defn}[Multi-Server Differential Privacy]
$\Pi$ is $(\eps,\delta)$-differentially private against $c$ corrupted servers if, for all $\Phi$ where $|C_s| \leq c$ and for every neighboring pair $\vec{x}\sim \vec{x}\,'$ differing on $i\notin C_u$,
$$
\view^{\Pi}_\Phi(\vec{x}) \approx_{\eps,\delta} \view^{\Pi}_\Phi(\vec{x}\,')
$$
\end{defn}

\begin{rem}
In the attacks we consider, the adversary's communications do not deviate from the protocol's specification. Borrowing language from cryptography, they are \emph{semi-honest} or \emph{passive}. We would naturally like protocols to ensure privacy against \emph{malicious} or \emph{active} adversaries, where messages are generated from arbitrary code. But our lower bounds hold even for the weaker family of protocols.
\end{rem}

Our work relies on a variety of constructions involving local randomizers, so we close this subsection with some  relevant notation. For any subset of servers $S$, let $R_{i,S}$ be the algorithm that, on input $x$, computes $(y_{i,1},\dots,y_{i,k}) \gets R_i(x)$ and reports only $\{y_{i,j}\}_{j\in S}$. For any event $E$ and disjoint subsets of servers $S, S'$, let $R_{i,S}(x) ~|~ R_{i,S'}(x)\in E$ denote the distribution of $\{y_{i,j}\}_{j\in S}$ conditioned on $\{y_{i,j}\}_{j\in S'} \in E$, where $\{y_{i,j}\}_{j\in[k]}$ are jointly drawn from $R_i(x)$. We use $R_{i,S}(\bD)$ and $R_{i,S}(\bD) ~|~ R_{i,S'}(\bD)\in E$ to denote the distributions when $x$ is first sampled from $\bD$.

\subsection{Online Algorithms}
\label{sec:online}

An online algorithm $M$ is specified by three algorithms $(M_\init,M_\update,M_\out)$. Algorithm \ref{alg:online} depicts how $M$ is executed on a stream $\vec{x}$ of length $n$: after initializing state, it repeatedly updates the state based upon the input stream.

\begin{algorithm}
\caption{The execution of an online algorithm $M= (M_\init,M_\update,M_\out)$}
\label{alg:online}

Initialize internal state $S_0\gets M_\init(\cdot)$

\For{$i\in [n]$}{
    Update internal state $S_i \gets M_\update(i,S_{i-1},x_i)$
}

Compute output $z_\out \gets M_\out(S_n)$

\Return{$z_\out$}
\end{algorithm}

To define privacy in this model, we mirror the previous section and define adversarial views. We make two assumptions: $M_\update$ is atomic and the privacy adversary can only view one internal state.

\begin{defn}[Internally Private Online Algorithms]
For any online algorithm $M$, input $\vec{x}$, and time of intrusion $t$, let $\view^M_t(\vec{x}) := S_t$ where $S_t$ is generated as in Algorithm \ref{alg:online}. $M$ is $(\eps,\delta)$-\emph{internally private} if and only if the following holds for every neighboring pair $\vec{x}\sim \vec{x}\,'$ and intrusion time $t$:
$$
\view^{M}_t(\vec{x}) \approx_{\eps,\delta} \view^{M}_t(\vec{x}\,')
$$
\end{defn}

The above definition originates in the thesis by Cheu \cite{Cheu21}. It is a relaxation of pan-privacy, wherein the adversary's view also includes the output $z_\out$ \cite{DNpRY10,AJM19,BCJM20,CU20}.

\section{From Multi-Server Protocols to Online Algorithms}
\label{sec:transform}


\begin{thm}
\label{thm:main}
Suppose $\Pi$ is a $k$-server protocol that takes $n$ inputs and offers $(\eps,\delta)$-privacy against $\lceil k/2\rceil$ corrupt servers. There exists a $(7\eps, O(e^{5\eps} \delta))$-internally-private online algorithm $A_\Pi$ that takes $m=O(e^{4\eps}n + e^{2\eps}\log (1/\beta))$ inputs with the following property: for any distribution $\bD$ over inputs,
$$
\sd{\Pi(\bD^n)}{A_\Pi(\bD^m) } \leq n\delta + \beta
$$
where $\bD^n$ is shorthand for $n$ i.i.d. samples from $\bD$.
\end{thm}

We proceed in three stages. First, we prove some essential technical lemmas regarding local randomizers. Next, we describe how to simulate $\Pi$ with an online algorithm in the case where $k=2$. Finally, we argue that any protocol with larger $k$ can be simulated by a two-server protocol with the same privacy parameters.

\subsection{Properties of Local Randomizers}
Although $\Pi$'s privacy guarantee does not imply any $R_i$ is differentially private, it is straightforward to show that any strict subset of the randomizer's outputs is $(\eps,\delta)$-private. 

\begin{clm}
\label{clm:subset-private}
For any user $i\in [n]$ and subset of servers $S\subset [k]$ where $|S| \leq \lceil k/2 \rceil$, $R_{i,S}$ is $(\eps,\delta)$-differentially private.
\end{clm}
\begin{proof}
Consider any attack $\Phi$ where $C_s=S$ and $i\notin C_u$. Fix any $\vec{x}\sim \vec{x}\,'$ that differ on $i$. Both $\view^\Pi_\Phi(\vec{x})$ and $\view^\Pi_\Phi(\vec{x}\,')$ contain messages from user $i$ to servers in $S$ in the same positions; closure under post-processing (Fact \ref{fact:post}) implies $R_{i,S}(x_i) \approx_{\eps,\delta} R_{i,S}(x'_i)$.
\end{proof}

It is easier to work with local randomizers that satisfy pure differential privacy than those that only satisfy approximate differential privacy. For this reason, we present the following technical lemma:

\begin{lem}
\label{lem:approx-to-pure}
If $R:\cX\to\cY$ is $(\eps,\delta)$-differentially private, there exists an algorithm $\tilde{R}$ that is $2\eps$-differentially private such that, for any $x\in\cX$, $\sd{R(x)}{\tilde{R}(x)}\leq \delta$.
\end{lem}

Proofs of Lemma \ref{lem:approx-to-pure} can be found in prior work; see e.g. Lemma 3.7 in Cheu and Ullman \cite{CU20}. By combining this lemma with Claim \ref{clm:subset-private}, we obtain a very useful corollary:

\begin{lem}
\label{lem:pure-subset}
For any user $i\in [n]$ and subset of servers $S\subset [k]$ where $|S| \leq \lceil k/2 \rceil$, there exists a $2\eps$-differentially private algorithm $\tilde{R}_{i,S}$ such that for any $x\in\cX$, $\sd{R_{i,S}(x)}{\tilde{R}_{i,S}(x)}\leq \delta$.
\end{lem}


\subsection{The Two-server Case}
\label{sec:two-server-transform}

\begin{thm}\label{thm:two-server-reduction}
Suppose $\Pi$ is a two-server protocol that takes $n$ inputs and offers $(\eps,\delta)$-privacy against one corrupt server. There exists a $(7\eps, O(e^{5\eps} \delta))$-internally-private online algorithm $A_\Pi$ that takes $m=O(e^{4\eps}n + e^{2\eps}\log (1/\beta))$ inputs with the following property: for any distribution $\bD$ over inputs,
$$
\sd{\Pi(\bD^n)}{A_\Pi(\bD^m) } \leq n\delta + \beta
$$
\end{thm}

We construct a sequence of algorithms $M_1,M_2,M_3$. Each approximates its predecessor and we show that $M_3$, by erasing unnecessary random variables, is our desired internally private algorithm $A_\Pi$.

\begin{rem}
$M_1$ and $M_3$ are online algorithms but to enhance readability, we avoid explicitly decomposing them into initialization, update, and output sub-routines as done in Section \ref{sec:online}.
\end{rem}

\subsubsection{Step One: Shifting to Pure Differential Privacy}

The pseudocode of $M_1$ is given in Algorithm \ref{alg:m1}. The sole difference between $M_1$ and the correct execution of $\Pi$ (Algorithm \ref{alg:multiserver}) is swapping $R_{i,1}$ with $\tilde{R}_{i,1}$, the $(2\eps,0)$-d.p. version of $R_{i,1}$. We do this to ease downstream analysis. Note that this step can be skipped if $\Pi$ already guarantees $\delta=0$.

\begin{algorithm}
    \caption{$M_1(\vec{x})$, an online algorithm that approximates $\Pi(\vec{x})$}
    \label{alg:m1}
    \For{$i \in [n]$}{
        $y_{i,1}\sim \tilde{R}_{i,1}(x_i)$ \tcc{Refer to Lemma \ref{lem:pure-subset}}
    
        $y_{i,2} \sim R_{i,2}(x_i)~|~R_{i,1}(x_i)=y_{i,1}$
    }
    
    $z_{1\to 2} \sim A_1(y_{1,1},\dots,y_{n,1})$
    
    $z_\out \gets A_2(z_{1\to 2}, y_{1,2}, \dots, y_{n,2})$
    
    \Return{$z_\out$}
\end{algorithm}

\begin{clm}
For any protocol inputs $\vec{x}$, $\sd{M_1(\vec{x})}{\Pi(\vec{x})}\leq n\delta$
\end{clm}
\begin{proof}
Lemma \ref{lem:pure-subset} implies that swapping out $R_{i,1}$ for $\tilde{R}_{i,1}$ changes the distribution only by $\delta$ at each of the $n$ sample points. A union bound completes the proof.
\end{proof}

\subsubsection{Step Two: Generating Messages to Server 2 via Bayesian Re-Sampling}

\begin{figure}[h]
    \centering
    \includegraphics[width=0.8\textwidth]{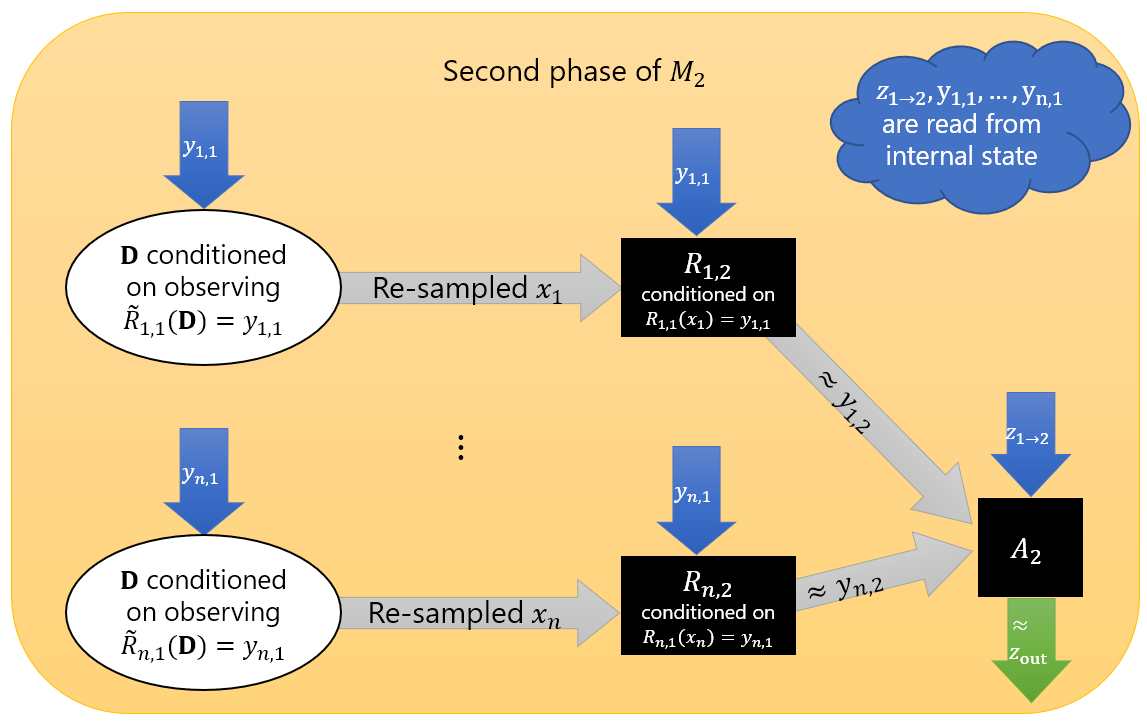}
    \caption{Visualization of how $M_2$ simulates the messages to server 2.}
    \label{fig:transform-2}
\end{figure}

The pseudocode of $M_2$ is given in Algorithm \ref{alg:m2}. It proceeds in two phases, each dedicated to simlating a server's inputs. Like $M_1$, it creates the messages to server 1 by running $\tilde{R}_{1,1}, \dots, \tilde{R}_{n,1}$ on the input. Unlike $M_1$, $M_2$ does not generate the messages to server 2 $\{y_{i,2}\}_{i\in [n]}$ directly from the input. Instead, it performs \emph{Bayesian re-sampling} as done by Joseph, Mao, Neel, and Roth \cite{JMNR19}: to produce $y_{i,2}$, it runs the local randomizer on a fresh sample from $\bD$ \emph{conditioned on having seen} $y_{i,1}$. Refer to Figure \ref{fig:transform-2} for a visualization of the second phase.

\begin{algorithm}
    \caption{$M_2(\vec{x})$, an algorithm that uses Bayesian re-sampling to simulate $M_1(\vec{x})$ when $\vec{x}\sim \bD^n$}
    \label{alg:m2}
    
    \tcc{First Phase: Create messages to server 1}
    
    \For{$i \in [n]$}{
        $y_{i,1} \sim \tilde{R}_{i,1}(x_i)$
    }
    
    $z_{1\to 2} \sim A_1(y_{1,1},\dots,y_{n,1})$

    \tcc{Second Phase: Create messages to server 2 by re-sampling user data}
    
    \For{$i \in [n]$}{
        $\hat{x}_i \sim \bD ~|~\tilde{R}_{i,1}(\bD)=y_{i,1}$
        
        $y_{i,2} \sim R_{i,2}(\hat{x}_i) ~|~ R_{i,1}(\hat{x}_i) = y_{i,1}$
    }

    $z_\out \gets A_2(z_{1\to 2}, y_{1,2}, \dots, y_{n,2})$

    \Return{$z_\out$}
\end{algorithm}

\begin{clm}
The distribution $M_2(\bD^n)$ is identical to $M_1(\bD^n)$
\end{clm}
\begin{proof}
Because the outputs of $M_1,M_2$ are determined by running $A_2$ on $(z_{1\to 2}, \{y_{i,2}\})_{i\in[n]})$, it will suffice to show that $(z_{1\to 2}, \{y_{i,2}\})_{i\in[n]})$ has the same distribution in both $M_2(\bD^n)$ and $M_1(\bD^n)$.

\begin{align*}
    &\pr{M_2(\bD^n)}{(z_{1\to 2}, \{y_{i,2}\}_{i\in[n]}) = (z,\vec{u})}\\
={}& \sum_{\vec{v}} \pr{M_2(\bD^n)}{(z_{1\to 2}, \{y_{i,2}\}_{i\in[n]}) = (z,\vec{u}),~ \forall i ~ y_{i,1} =v_i} \\
={}& \sum_{\vec{v}} \pr{M_2(\bD^n)}{(z_{1\to 2}, \{y_{i,2}\}_{i\in[n]}) = (z,\vec{u})~|~ \forall i ~ y_{i,1} =v_i} \cdot \pr{M_2(\bD^n)}{\forall i ~ y_{i,1} =v_i} \\
={}& \sum_{\vec{v}} \pr{M_2(\bD^n)}{(z_{1\to 2}, \{y_{i,2}\}_{i\in[n]}) = (z,\vec{u})~|~ \forall i ~ y_{i,1} =v_i} \cdot \pr{y_{i,1} \sim \tilde{R}_{i,1}(\bD)}{\forall i ~ y_{i,1} =v_i} \tag{By construction}\\
={}& \sum_{\vec{v}} \pr{}{A_1(\vec{v})=z} \cdot \prod_{i=1}^n \underbrace{\pr{M_2(\bD^n)}{y_{i,2} =u_i ~|~  y_{i,1} =v_i}}_{T} \cdot \pr{y_{i,1} \sim \tilde{R}_{i,1}(\bD)}{\forall i ~ y_{i,1} =v_i} \stepcounter{equation} \tag{\theequation} \label{eq:term-t}
\end{align*}
\eqref{eq:term-t} follows from the fact that $y_{1,2},\dots,y_{n,2}$ are mutually independent and also independent of $z_{1\to 2}$.

We know that the term $T$ is equal to $\pr{M_1(\bD^n)}{ y_{i,2} =u_i ~|~y_{i,1} = v_i}$ because $M_2$ merely changes the sampling order from ``data, message 1, message 2'' to ``message 1, data, message 2.'' Thus,
\begin{align*}
\eqref{eq:term-t} &= \sum_{\vec{v}} \pr{}{A_1(\vec{v})=z} \cdot \pr{M_2(\bD^n)}{\forall i ~ y_{i,2} =u_i ~|~ \forall i ~ y_{i,1} =v_i} \cdot \pr{y_{i,1} \sim \tilde{R}_{i,1}(\bD)}{\forall i ~ y_{i,1} =v_i} \\
    &= \sum_{\vec{v}} \pr{M_1(\bD^n)}{(z_{1\to 2}, \{y_{i,2}\}_{i\in[n]}) = (z,\vec{u})~|~ \forall i ~ y_{i,1} =v_i} \cdot \pr{M_1(\bD^n)}{\forall i ~ y_{i,1} =v_i}\\
    &= \pr{M_1(\bD^n)}{(z_{1\to 2}, \{y_{i,2}\}_{i\in[n]}) = (z,\vec{u})} \qedhere
\end{align*}
\end{proof}


\subsubsection{Step Three: Implementing Bayesian Re-Sampling via Rejection Sampling}

\begin{figure}
    \centering
    \includegraphics[width=0.9\textwidth]{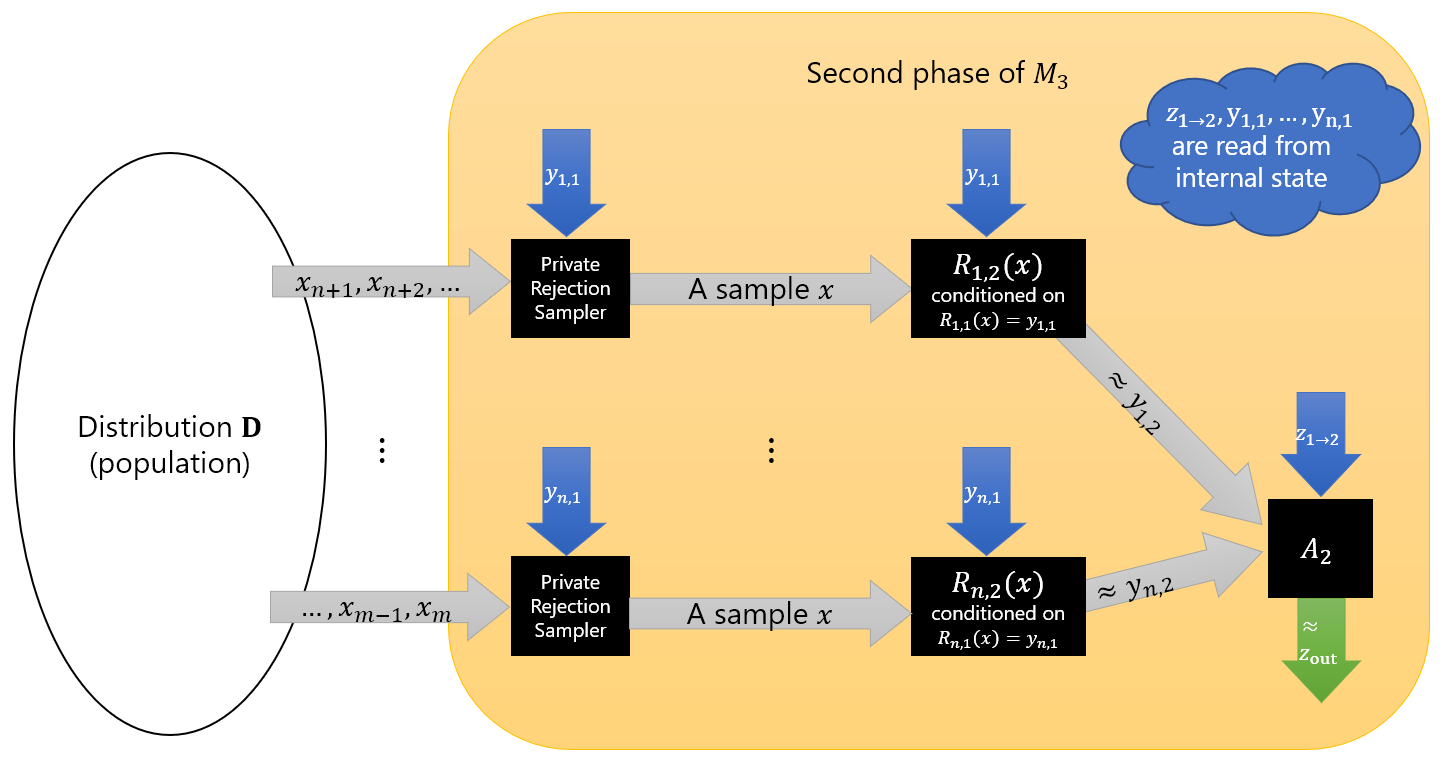}
    \caption{Visualization of how $M_3$ simulates the messages to server 2, assuming it has sample access to $\bD$.}
    \label{fig:transform-3}
\end{figure}

$M_2$ requires us to sample from $\bD$ conditioned on $\tilde{R}_{i,1}(\bD)=y_{i,1}$, for every $i$. Since $\bD$ is the unknown distribution that is the subject of study, $M_2$ can only be a thought-experiment. But we approximate the desired conditional distribution by performing private \emph{rejection sampling} on independent samples from $\bD$, as done by Joseph et al. \cite{JMNR19}. This modification is presented in $M_3$ (Algorithm \ref{alg:m3}). It takes in $m > n$ samples, the excess being used for the rejection sampling. The updated second phase is visualized in Figure \ref{fig:transform-3}.

\begin{algorithm}
    \caption{$M_3(\bD^m)$, an online algorithm that approximates $M_2(\bD^n)$}
    \label{alg:m3}
    
    \tcc{First Phase: Create messages to server 1}
    \For{$i \in [n]$}{
        
        $y_{i,1} \sim \tilde{R}_{i,1}(x_i)$
    }
    
    $z_{1\to 2} \sim A_1(y_{1,1},\dots,y_{n,1})$
    
    \tcc{Second Phase: Create messages to server 2 by re-sampling user data}
    \tcc{Re-sampling approximated by rejection sampling}
    
    $i\gets 1$
    
    \For{$h \in \{n+1,\dots, m\}$}{
        Compute acceptance rate $\textit{rate}_h \gets \frac{\pr{}{\tilde{R}_{i,1}(x_h) = y_{i,1}}}{2 \cdot \max_u \pr{}{\tilde{R}_{i,1}(u) = y_{i,1}}}$
        
        $a_h \sim \Ber(\textit{rate}_h)$
        
        Erase $\textit{rate}_h$ from internal state
        
        \If{$a_h=1$}{
        
            $y_{i,2} \sim R_{i,2}(x_h) ~|~ R_{i,1}(x_h) = y_{i,1}$
            
            Erase $y_{i,1}$ from internal state
            
            $i\gets i+1$
            
            \If{$i=n+1$}{Break loop}
        }
        
    }
    
    $z_\out \gets A_2(z_{1\to 2}, y_{1,2}, \dots, y_{n,2})$
    
    \Return{$z_\out$}
\end{algorithm}

\begin{clm}
For any $\beta\in(0,1)$, there is some $m=O(e^{4\eps}n + e^{2\eps}\log (1/\beta))$ where $\sd{M_3(\bD^m)}{M_2(\bD^n)}\leq \beta$
\end{clm}
\begin{proof}
As before, our analysis will condition on an arbitrary realization of $(z_{1\to 2}, \{y_{i,1}\})_{i\in[n]})$.

We will in fact spend much of the proof studying $M_3^\infty(\bD^\infty)$, a version of $M_3$ which takes in an \emph{unbounded} stream of samples (replace the second for-loop with a while-loop). After establishing basic facts, we show that this alternate algorithm correctly simulates $M_2(\bD^n)$. Then we show that this alternate algorithm consumes only $m$ samples with $\geq 1-\beta$ probability. Therefore, stopping at the $m$-th sample changes the overall distribution by at most $\beta$.

\textit{Facts about $M_3^\infty(\bD^\infty)$}: Let $\eta_1$ be the number of iterations of the while-loop until we obtain $y_{1,2}$ and move pointer $i$ to 2. For $i>1$, let $\eta_i$ be number of iterations between sampling $y_{i-1,2}$ and sampling $y_{i,2}$.

We now characterize the distribution of every $\eta_i$. To do so, note that, for any step $h$ in the while-loop before $y_{i,2}$ is sampled,
\begin{align*}
    \pr{x_h\sim \bD}{a_h=1} &= \int_{v=0}^1 v\cdot \pr{x_h \sim \bD}{\textit{rate}_h=v} ~dv\\
    &= \int_{v=0}^1 v\cdot \pr{x_h \sim \bD}{\frac{\pr{}{\tilde{R}_{i,1}(x) = y_{i,1}}}{2 \cdot \max_u \pr{}{\tilde{R}_{i,1}(u) = y_{i,1}}} = v} ~dv\\
    &=: p_i \in \left[ \frac{1}{2e^{2\eps}} ,~ \half \right] \stepcounter{equation} \tag{\theequation} \label{eq:accept-rate}
\end{align*}
The last step comes from Lemma \ref{lem:pure-subset}. The immediate corollary is that $\eta_i$ is drawn from $\Geo(p_i)$, the geometric distribution characterizing the number of $\Ber(p_i)$ trials until success.

\textit{Correctness of $M_3^\infty(\bD^\infty)$}: Because we have shown that the acceptance rate of a sample is nonzero ($p_i>0$), the algorithm \emph{eventually} samples $y_{i,2}$ for every $i$. We claim this implies correct simulation. Specifically, conditioned on $a_h=1$, we claim that $x_h$ is drawn from the correct posterior $\bD ~|~ \tilde{R}_{i,1}(\bD) y_{i,1}$. The calculation below is adapted from an equivalent step in Joseph et al. \cite{JMNR19}:

\begin{align*}
\pr{}{x_h = x~|~ a_h=1} &= \pr{}{a_h=1 ~|~ x_h = x}\cdot \frac{\pr{}{x_h = x}}{\pr{}{a_h=1}}\\
    &= \frac{\pr{}{\tilde{R}_{i,1}(x) = y_{i,1}}}{2 \cdot \max_u \pr{}{\tilde{R}_{i,1}(u) = y_{i,1}}} \cdot \frac{\pr{}{x_h = x}}{\sum_{\hat{x}} \pr{}{x_h = \hat{x}} \cdot \frac{\pr{}{\tilde{R}_{i,1}(\hat{x})=y_{i,1}}}{2 \cdot \max_u \pr{}{\tilde{R}_{i,1}(u) = y_{i,1}}}}\\
    &= \frac{\pr{}{\tilde{R}_{i,1}(x) = y_{i,1}}\cdot \pr{}{x_h = x}}{\sum_{\hat{x}} \pr{}{x_h = \hat{x}} \cdot \pr{}{\tilde{R}_{i,1}(\hat{x})=y_{i,1}}} \\
    &= \pr{v\sim \bD}{v=x~|~\tilde{R}_{i,1}(v)=y_{i,1}}
\end{align*}

\textit{Distance between $M_3^\infty(\bD^\infty)$ and $M_3(\bD^m)$}: The total number of samples consumed by $M_3^\infty(\bD^\infty)$ is $n+\sum_{i=1}^n \eta_i$. Since we know each $\eta_i$ is a geometric random variable, the following lemma is useful:
\begin{lem}[Tail Bound for Geometric Convolutions]
\label{lem:geo-chernoff}
Fix any $p_1,\dots,p_n, \ell \in (0,1/2]$ such that $\ell \leq p_i$ for every $i\in[n]$. If we sample $\eta_i$ from $\Geo(p_i)$ for every $i\in [n]$, then
$$
\sum_{i=1}^n \eta_i = O\paren{ \frac{n}{\ell}\ln \frac{1}{\ell} + \frac{1}{\ell}\ln \frac{1}{\beta} }
$$
with probability at least $1-\beta$, for any $\beta \in (0,1)$.
\end{lem}
The proof can be found in Appendix \ref{apdx:deferred}. From \eqref{eq:accept-rate}, we have that $\ell=1/2e^{2\eps}$ so that substitution implies
\begin{align*}
\sum_{i=1}^n \eta_i &= O\paren{ e^{2\eps}n \ln (2e^{2\eps}) + e^{2\eps}\ln \frac{1}{\beta} }\\
    &= O\paren{ e^{4\eps}n + e^{2\eps}\ln \frac{1}{\beta} } \tag{$\ln(2e^{2\eps})< 1+2\eps$}
\end{align*}
So setting $m$ to the above (plus $n$) ensures $M_3(\bD^m)$ is within $\beta$ of $M_3^\infty(\bD^\infty)$ in statistical distance. Because we know $M_3^\infty(\bD^\infty)$ matches $M_2(\bD^n)$, the proof is complete.
\end{proof}

\subsubsection{Proving Internal Privacy}
We finally prove $M_3$ ensures internal privacy. The guarantee comes from an invariant that the algorithm maintains: for any user data $x_i$, $M_3$ keeps at most one message generated from $x_i$ inside the internal state. If the user belongs to the first batch (processed by the first for-loop), we are guaranteed that the message is generated by some algorithm $\tilde{R}_{i,1}$ which Lemma \ref{lem:pure-subset} ensures is private. Otherwise, we must deal with some technicalities regarding the rejection sampling. But we are able to show that the user's message is close in distribution to $R_{i,2}(x_i)$. Claim \ref{clm:subset-private} ensures this is close to $R_{i,2}(x'_i)$.

We formally state our claim below:
\begin{clm}
\label{clm:transform-privacy}
If $\Pi$ is $(\eps,\delta)$-private against 1 corrupt server, then $M_3$ is $(7\eps,O(e^{5\eps}\delta)$-internally-private.
\end{clm}

The claim will follow from case analysis. We state these cases as separate sub-claims.

\newcommand{\hi}{\hat{\imath}}

\begin{clm}
\label{clm:privacy-1}
If $\Pi$ is $(\eps,\delta)$-private against 1 corrupt server and $\vec{x},\vec{x}\,'$ differ only on $i^* \in [n]$, then for any intrusion time $t>i^*$ $$\view^{M_3}_t(\vec{x}) \approx_{2\eps} \view^{M_3}_t(\vec{x}\,')$$
\end{clm}

\begin{proof}
First, suppose $t\in [n]$. Here, the adversary intrudes before the second for-loop. The state it obtains consists of all random variables created by $M_3(\vec{x})$ (resp. $M_3(\vec{x}\,')$) so far, which includes the messages $\{y_{i,1}\}$ (resp. $\{y'_{i,1}\}$) for $i\in [t]$. When $t=n$, the state also includes $z_{1\to 2}$ (resp. $z'_{1\to 2}$) due to the way we define online algorithms; see Algorithm \ref{alg:online}. But these random variables are obtained from post-processing the messages. Thus, it will suffice to show $$\{y_{i,1}\}_{i\in [n]} \approx_{2\eps} \{y'_{i,1}\}_{i\in [n]}$$ Because $\tilde{R}_{i^*,1}$ is $(2\eps,0)$-private, we have that $y_{i^*,1} \approx_{2\eps} y'_{i^*,1}$. And by construction, every other $y_{i,1}$ is identically distributed with $y'_{i,1}$. The claim follows by the mutual independence of the messages.

For other intrusion times $t \in [n+1,m]$, observe that $x_{i^*}$ (resp. $x'_{i^*}$) is never read again so that the claim again holds by post-processing. 
\end{proof}

\begin{clm}
\label{clm:privacy-2}
If $\Pi$ is $(\eps,\delta)$-private against 1 corrupt server and $\vec{x},\vec{x}\,'$ that differ only on $i^*\in[n+1,m]$, then for any intrusion time $t >i^*$, $$\view^{M_3}_t(\vec{x}) \approx_{7\eps,3e^{5\eps}\delta} \view^{M_3}_t(\vec{x}\,')$$
\end{clm}
\begin{proof}
In this case, notice that the view on input $\vec{x}$ is\footnote{Similar to before, the views include $z_\out$ and $z'_\out$ when $t=m$ but closure under post-processing will again ensure that this does not affect the proof.}
$$
\view^{M_3}_t(\vec{x}) = (z_{1\to 2}, y_{1,2},\dots, y_{\hi-1,2}, y_{\hi,1}, \dots, y_{n,1}, a_{n+1},\dots,a_t)
$$
where each $a_h$ is a bit sampled from $\Ber(\textit{rate}_h)$ and $\hi = 1+\sum_{h=n+1}^t a_h$ is the index of the user in the first batch whose data we are re-sampling. Likewise,
$$
\view^{M_3}_t(\vec{x}\,') = (z'_{1\to 2}, y'_{1,2},\dots, y'_{\hi'-1,2},y'_{\hi',1}, \dots, y'_{n,1}, a'_{n+1},\dots,a'_t)
$$
where each $a'_h$ is sampled from $\Ber(\textit{rate}'_h)$ and $\hi' = 1+\sum_{h=n+1}^t a'_h$.

Let $v(i^*) \in [n]$  (resp. $v'(i^*)$) be the value of the pointer $i$ at the beginning of iteration $i^*$ in $M_3(\vec{x})$ (resp. $M_3(\vec{x})\,'$).

We will argue that $(a_{i^*},y_{v(i^*),2}) \approx_{7\eps,3e^{5\eps}\delta} (a'_{i^*},y_{v'(i^*),2})$. This suffices because the other pairs of variables in $\view^{M_3}_t(\vec{x})$ are either (a) identically distributed with counterparts in $\view^{M_3}_t(\vec{x}\,')$ or (b) obtained by post-processing $(a_{i^*},y_{v(i^*),2})$.

By the privacy of every $\tilde{R}_{i,1}$, we have that $\pr{}{a_h=1},\pr{}{a'_h=1}\in [1/2e^{2\eps},1/2]$ for any $h\in [n+1,m]$. Then,
\begin{align*}
\frac{\pr{}{a_h = 1}}{\pr{}{a'_h=1}} &\leq \frac{1/2}{1/2e^{2\eps}} = e^{2\eps}\\
\frac{\pr{}{a_h = 0}}{\pr{}{a'_h=0}} &\leq \frac{1 - 1/2e^{2\eps}}{1/2} = 2-e^{-2\eps} < e^{2\eps} \tag{Taylor series}
\end{align*}


Consequently,
\begin{align*}
    & \pr{}{(a_{i^*},y_{v(i^*),2}) \in E}\\
    ={}& \pr{}{(a_{i^*},y_{v(i^*),2}) \in E ,~ a_{i^*} =0} + \pr{}{(a_{i^*},y_{v(i^*),2}) \in E ,~ a_{i^*} =1} \\
    \leq{}& e^{2\eps}\cdot \pr{}{(a_{i^*},y_{v(i^*),2}) \in E ~|~ a_{i^*} =0} \cdot \pr{}{a'_{i^*}=0} \\
    &+ e^{2\eps}\cdot \pr{}{(a_{i^*},y_{v(i^*),2}) \in E ~|~ a_{i^*} =1} \cdot \pr{}{a'_{i^*}=1} \\
    ={}& e^{2\eps}\cdot \pr{}{(a'_{i^*},y'_{v'(i^*),2}) \in E ~|~ a'_{i^*} =0} \cdot \pr{}{a'_{i^*}=0} \\
    &+ e^{2\eps}\cdot \pr{}{(a_{i^*},y_{v(i^*),2}) \in E ~|~ a_{i^*} =1} \cdot \pr{}{a'_{i^*}=1} \stepcounter{equation} \tag{\theequation} \label{eq:privacy-2}
\end{align*}
The last step comes from two facts. First, $x_{i^*}$ (resp. $x'_{i^*}$) will not be used in the creation of $y_{v(i^*),2}$ (resp. $y'_{v(i^*),2}$) when $a_{i^*}=0$ (resp. $a_{i^*}'=0$) because the bit indicates rejection. Second, $v(i^*)$ is identically distributed with $v'(i^*)$ because the inputs $\vec{x},\vec{x}\,'$ are by definition identical prior to $i^*$.

If $a_{i^*}=1$, $x_{i^*}$ will be used to generate $y_{v(i^*),2}$. Let $\cY_j$ be the range of $R_{v(i^*),j}$ and we assume without loss of generality it is discrete.
\begin{align*}
&\pr{}{(a_{i^*},y_{v(i^*),2}) \in E ~|~ a_{i^*} =1}\\
={}& \sum_{y \in \cY_1} \pr{}{(a_{i^*},y_{v(i^*),2}) \in E ~|~  y_{v(i^*),1} = y,~ a_{i^*} =1} \cdot \pr{}{y_{v(i^*),1} = y ~|~ a_{i^*} =1}\\
={}& \sum_{y \in \cY_1} \pr{}{(a_{i^*},y_{v(i^*),2}) \in E ~|~ y_{v(i^*),1} = y,~ a_{i^*} =1} \cdot \pr{}{\tilde{R}_{v(i^*),1}(x_{v(i^*)}) = y}\\
\leq{}& e^{2\eps} \cdot \sum_{y \in \cY_1} \pr{}{(a_{i^*},y_{v(i^*),2}) \in E ~|~ y_{v(i^*),1} = y,~ a_{i^*} =1} \cdot \pr{}{\tilde{R}_{v(i^*),1}(x_{i^*}) = y} \tag{Lemma \ref{lem:pure-subset}}\\
\leq{}& e^{2\eps} \cdot \paren{ \delta+ \sum_{y \in \cY_1} \pr{}{(a_{i^*},y_{v(i^*),2}) \in E ~|~ y_{v(i^*),1} = y,~ a_{i^*} =1} \cdot \pr{}{R_{v(i^*),1}(x_{i^*}) = y}} \tag{Lemma \ref{lem:pure-subset}}\\
={}&  e^{2\eps} \cdot \pr{}{R_{v(i^*),2}(x_{i^*}) \in F} + e^{2\eps}\delta \stepcounter{equation} \tag{\theequation} \label{eq:privacy-2a}
\end{align*}
where $F$ be the subset of $\cY_2$ such that $y\in F$ if and only if $(1,y)\in E$.

Symmetric reasoning yields
\begin{equation}
\label{eq:privacy-2b}
\pr{}{(a'_{i^*},y'_{v'(i^*),2}) \in E ~|~ a'_{i^*} =1} \geq e^{-2\eps}\cdot \pr{}{R_{v'(i^*),2}(x'_{i^*}) \in F} -e^{-2\eps}\delta
\end{equation}
We also know that $v(i^*)$ is identically distributed with $v'(i^*)$ and, for any $i$,
\begin{equation}
\label{eq:privacy-2c}
\pr{}{R_{i,2}(x_{i^*}) \in F} \leq e^\eps \pr{}{R_{i,2}(x'_{i^*}) \in F} + \delta,
\end{equation}
Combining \eqref{eq:privacy-2a}, \eqref{eq:privacy-2b}, and \eqref{eq:privacy-2c}, we obtain
$$
\pr{}{(a_{i^*},y_{v(i^*),2}) \in E ~|~ a_{i^*} =1} < e^{5\eps}\cdot  \pr{}{(a'_{i^*},y'_{v'(i^*),2}) \in E ~|~ a'_{i^*} =1} + 3e^{3\eps}\delta
$$

When we substitute the above into \eqref{eq:privacy-2}, we have
\begin{align*}
    & \pr{}{(a_{i^*},y_{v(i^*),2}) \in E}\\
    <{}& e^{2\eps}\cdot \pr{}{(a'_{i^*},y'_{v'(i^*),2}) \in E ~|~ a'_{i^*} =0} \cdot \pr{}{a'_{i^*}=0} \\
    &+ e^{2\eps}\cdot \paren{e^{5\eps}\cdot  \pr{}{(a'_{i^*},y'_{v'(i^*),2}) \in E ~|~ a'_{i^*} =1} + 3e^{3\eps}\delta} \cdot \pr{}{a'_{i^*}=1}\\
    <{}& e^{7\eps}\cdot \pr{}{(a'_{i^*},y'_{v'(i^*),2}) \in E} + 3e^{5\eps}\delta
\end{align*}
which is what we wanted to prove.
\end{proof}

\paragraph{Why are Interactive Protocols Difficult to Transform?}
One can imagine a two-server protocol where server 2 sends some $z_{2\to 1}$ to server 1, who then produces the output $z_\out$. This is challenging to transform into a private online algorithm. To see why, recall that we need to simulate the view of the server who produces output, which is here the joint random variable $(y_{1,1},\dots,y_{n,1},z_{1\to 2},z_{2\to 1})$. Our method allows us to privately simulate a different random variable $(y_{1,2},\dots,y_{n,2},z_{1\to 2},z_{2\to 1})$, where $z_{2\to 1}$ is the message server 2 produces for server 1. We could attempt to iteratively replace each $y_{i,2}$ with $y_{i,1}$ by again using Bayesian re-sampling, but the construction must now involve $z_{1\to 2}$. This random variable is obtained from $n$ independent samples from $\bD$, unlike one in Algorithm \ref{alg:m2}. Rejection sampling is now quite difficult: we would need a way to compute an acceptance rate for a batch of $n$ users, all while trying to maintain privacy of the internal state after reading each user's data.

\subsection{The Multi-server Case}

\begin{figure}[h]
    \centering
    \includegraphics[width=0.9\textwidth]{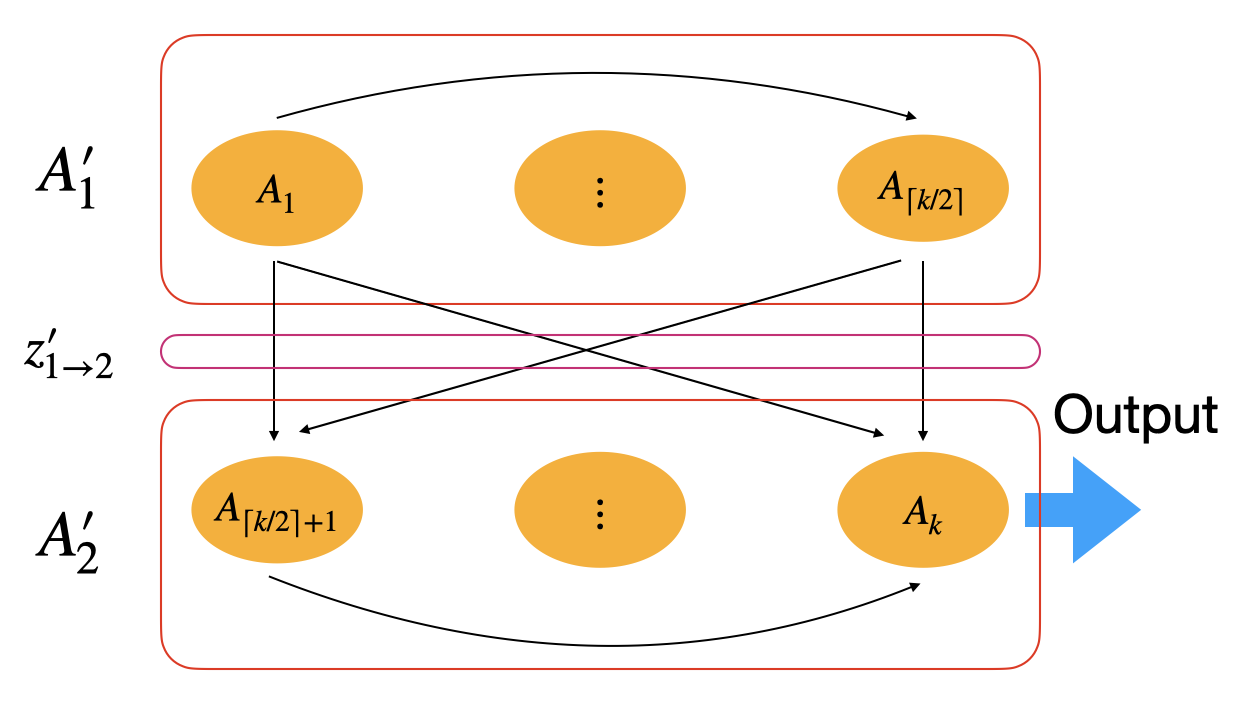}
    \caption{Visualization of how to use two servers to simulate the execution of $k$ servers.}
    \label{fig:reduction-1}
\end{figure}

We construct a reduction from the $k$-server case to the two-server case. We use the first server in the two-server protocol to simulate first half of servers in the $k$-server protocol and use the second server simulate the second half (see Figure \ref{fig:reduction-1}).

\begin{clm} \label{clm:k-server-reduction}
If the $k$-server protocol $\Pi=(\{R_i\}_{i\in[n]},\{A_j\}_{j\in[k]},G)$ is $(\eps,\delta)$-differentially private against $\lceil k/2\rceil$ corrupted servers, then there exists a two-server protocol $\Pi'=(\{R'_i\}_{i\in[n]},\{A'_j\}_{j\in[2]},G')$ which is $(\eps,\delta)$-differentially private against 1 corrupted server and $\Pi(\vec{x})=\Pi'(\vec{x})$ for any input $\vec{x}=(x_1,\dots,x_n)$.
\end{clm}

\begin{proof}
Recall that the $i$-th local randomizer in $\Pi$ has the form $R_i(x_i)=(y_{i,1},\dots,y_{i,k})$. $R'_i(x_i)$ (Algorithm \ref{alg:R'}) constructs $(y'_{i,1},y'_{i,2})$, where the first element is $(y_{i,1},\dots,y_{i,\lceil k/2\rceil})$ and the second is $(y_{i,\lceil k/2\rceil+1},\dots,y_{i,k})$.

\begin{algorithm}
    \SetAlgoLined
    \LinesNumbered
    \caption{$R'_i$, the $i$-th local randomizer in the two-servers protocol\label{alg:R'}}
    Let $R_i$ be the $i$-th local randomizer in the $k$-server protocol.
    
    \KwIn{$x_i$}
    Sample $(y_{i,1},\dots,y_{i,k})\sim R_i(x_i)$.
    
    Let $y'_{i,1}=(y_{i,1},\dots,y_{i,\lceil k/2\rceil})$ and $y'_{i,2}=(y_{i,\lceil k/2\rceil+1},\dots,y_{i,k})$.
    
    \Return{$(y'_{i,1},y'_{i,2})$}
\end{algorithm}

$A'_1$ (Algorithm \ref{alg:A'_1}) simulates the execution of $A_1\dots A_{\lceil k/2\rceil}$ and $A'_2$ (Algorithm \ref{alg:A'_2}) simulates $A_{\lceil k/2\rceil+1}\dots A_k$. Since we assume that all servers in $\Pi$ are named according to a topological ordering, $A'_2$ will not produce a message destined for a server simulated by $A'_1$: there is no interaction between the two servers. Let $z'_{1\to 2}$ be the collection of messages from the first half of the $k$ servers to the second half.

\begin{algorithm}
    \SetAlgoLined
    \LinesNumbered
    \caption{$A'_1$, the first server in the two-server protocol\label{alg:A'_1}}
    Let $A_1,\dots,A_{\lceil k/2\rceil}$ be the $1$-st to $\lceil k/2\rceil$-th servers in the $k$-server protocol.
    
    \KwIn{$y'_{1, 1},\dots,y'_{n, 1}$, where $y'_{i,1}=(y_{i,1},\dots,y_{i,\lceil k/2\rceil+1}$) for $i\in[n]$}
    
    Sample $(z_{1\to2},\dots,z_{1\to k})\sim A_1(y_{1,1},\dots,y_{n,1})$.
    
    \For{$j\in\{2,\dots,\lceil k/2\rceil\}$}{
        Sample $(z_{j\to j+1},\dots,z_{j\to k})\sim A_j(y_{1,j},\dots,y_{n,j},z_{1\to j},\dots,z_{j-1\to j})$.
    }
    
    Let $z'_{1\to 2}=(z_{1\to \lceil k/2\rceil+1},\dots, z_{1\to k},\dots,z_{\lceil k/2\rceil\to \lceil k/2\rceil+1},\dots,z_{\lceil k/2\rceil\to k})$.
    
    \Return{$z'_{1\to2}$}

\end{algorithm}

\begin{algorithm}
    \SetAlgoLined
    \LinesNumbered
    \caption{$A'_2$, the second server in the two-server protocol\label{alg:A'_2}}
    Let $A_{\lceil k/2\rceil+1},\dots,A_k$ be the $\lceil k/2\rceil+1$-th to $k$-th servers in the $k$-server protocol.
    
    \KwIn{$y'_{1, 2},\dots,y'_{n, 2},z'_{1\to2}$, where $y'_{i,2}=(y_{i,\lceil k/2\rceil+1},\dots,y_{i,k}$) for $i\in[n]$ and  $z'_{1\to 2}=(z_{1\to \lceil k/2\rceil+1},\dots, z_{1\to k},\dots,z_{\lceil k/2\rceil\to \lceil k/2\rceil+1},\dots,z_{\lceil k/2\rceil\to k})$}
    
    \For{$j\in\{\lceil k/2\rceil+1,\dots,k-1\}$}{
        Sample $(z_{j\to j+1},\dots,z_{j\to k})\sim A_j(y_{1,j},\dots,y_{n,j},z_{1\to j},\dots,z_{j-1\to j})$.
    }
    
    Sample $z_{out}\sim A_k(y'_{1\to k},\dots,y'_{n\to k},z_{1\to k},\dots, z_{k-1\to k}$
    
    \Return{$z_{out}$}
\end{algorithm}

The new protocol preserves accuracy because the output of $A'_2$ is identical to the output of $A_k$.

We now argue that $\Pi'$ is $(\eps,\delta)$-differentially private against 1 corrupted server if $\Pi$ is $(\eps,\delta)$-differentially private against $\lceil k/2\rceil$ corrupted servers. This is done by arguing any attack $\Phi'=(C'_u,C'_s)$ against $\Pi'$ corresponds to an attack $\Phi$ against $\Pi$. Specifically, let the set of corrupted users in $\Phi$ to be $C_u=C'_u$. If the set of corrupted servers in $\Phi'$ is $C'_s=\{1\}$, let $C_s=\{1,\dots,\lceil k/2 \rceil\}$. If $C'_u=\{2\}$, let $C_s=\{\lceil k/2 \rceil+1,\dots,k\}$. Note that $|C_s|\leq\lceil k/2\rceil$ and the view of the adversary in both attacks are identical. Since $\Pi$ is $(\eps,\delta)$-differentially private against $\Phi$, $\Pi'$ must be $(\eps,\delta)$-differentially private against $\Phi'$.
\end{proof}

Combining Theorem \ref{thm:two-server-reduction} and Claim \ref{clm:k-server-reduction}, we finally arrive at Theorem \ref{thm:main}. We restate it below for convenience:

\begin{thm*}[Copy of Theorem \ref{thm:main}]
Suppose $\Pi$ is a $k$-server protocol that takes $n$ inputs and offers $(\eps,\delta)$-privacy against $\lceil k/2\rceil$ corrupt servers. There exists a $(7\eps, O(e^{5\eps} \delta))$-internally-private algorithm $A_\Pi$ that takes $m=O(e^{4\eps}n + e^{2\eps}\log (1/\beta))$ inputs with the following property: for any distribution $\bD$ over inputs,
$$
\sd{\Pi(\bD^n)}{A_\Pi(\bD^m) } \leq n\delta + \beta
$$
\end{thm*}

\section{Lower Bounds for Multi-Server Protocols}
\label{sec:lower-bounds}

We are ready to invoke lower bounds proved by Cheu \& Ullman \cite{CU20} and Amin, Joseph, and Mao \cite{AJM19}. As previously mentioned, these results were developed for pan-privacy---where an adversary's view includes one internal state and the output of the protocol---but the arguments only require privacy of the internal state. Hence, we restate the theorems in terms of internal privacy, as done by Cheu \cite{Cheu21}. The definitions are also taken from that work.

To streamline the presentation, we assume $\eps=O(1)$ and $\delta \ll 1/n$.

\subsection{Parity Learning}
\newcommand{\err}{\mathrm{err}}
Let $\cX = \{\pm 1\}^{d+1}$ be the domain; we will treat the last bit of every member string as the label of the string. The error of a parity function $(\ell,b) \in 2^{[d]}\times \{ \pm 1 \}$ with respect to a distribution $\bD$ over $\cX$ is $$
\err_\bD(\ell,b):=\pr{x\sim \bD}{b\cdot \prod_{j\in \ell} x_j \neq x_{d+1}}
$$

\begin{defn}[Parity Learning]
An algorithm $M$ performs $(d,t,\alpha)$-parity learning with sample complexity $n$ if it takes $n$ independent samples from a distribution $\bD$ over $\cX$ and reports a tuple $(L,B) \in 2^{[d]}\times \{ \pm 1 \}$ such that, with probability 99/100, $|L|\leq t$ and
$$
\err_\bD(L,B) \leq \min_{\ell,b}\err_\bD(\ell,b) +\alpha
$$
\end{defn}

The following theorem can be found in \cite{Cheu21}. $\binom{d}{\leq t}$ is shorthand for $\sum_{j=0}^t \binom{d}{j}$.

\begin{thm}
\label{thm:internalDP-parity}
If online algorithm $M$ performs $(d,t,\alpha)$-parity learning with sample complexity $n$ and is $(\eps,\delta)$-internally private for $\delta=0$ or $\delta\log \paren{ \binom{d}{\leq t}/\delta} \ll \alpha^2\eps^2/\binom{d}{\leq t}$, then $n= \Omega\paren{\sqrt{\binom{d}{\leq t}}/\alpha\eps}$.
\end{thm}

By combining the above with Theorem \ref{thm:main}, we arrive at the following bound on the sample complexity of any parity learner in the (non-interactive) multi-server model.

\begin{thm}
\label{thm:multiserver-parity}
If $\Pi$ is a $k$-server protocol that solves $(d,t,\alpha)$-parity learning with sample complexity $n$ and offers $(\eps,\delta)$-differential privacy against $\lceil k/2\rceil$ corrupt servers for $\delta=0$ or $\delta\log \paren{ \binom{d}{\leq t}/\delta} \ll \alpha^2\eps^2/\binom{d}{\leq t}$, then $n=\Omega\paren{\sqrt{\binom{d}{\leq t}}/\alpha\eps}$
\end{thm}

In \cite{KLNRS08}, Kasiviswanathan et al. show that just $O(d/\alpha\eps)$ samples suffice under $\eps$-central privacy.

\begin{proof}[Proof of Theorem \ref{thm:multiserver-parity}]
Fix $\beta$ to be a sufficiently small constant, e.g. $10^{-5}$. From Theorem \ref{thm:main} and our bound on $\eps$, there must be a $(O(\eps), O(\delta))$-internally private algorithm which solves $(d,t,\alpha)$-parity learning with asymptotically identical sample complexity. The failure probability of this internally private learner differs from the $k$-server learner by at most $n\delta +\beta$. Note that this is bounded by $10^{-4}$ due to the magnitude of $\delta$ and our choice of $\beta$. Moreover, the proof of Theorem \ref{thm:internalDP-parity} is flexible enough to accommodate that small change in failure probability; the lower bound carries over.
\end{proof}

The proofs for the other theorems in this section are virtually identical, so we omit them for brevity.

\subsection{Feature Selection}
\begin{defn}[Feature Selection Problem]
Let $\alpha$ be any real in the interval $(0,1/2)$ and let $d$ be any integer larger than 1. An algorithm $M$ \emph{solves $(\alpha,d)$-feature selection with sample complexity $n$} if, for any distribution $\bD$ over $\zo^d$, it takes $n$ independent samples from $\bD$ and selects a coordinate $J \in [d]$ such that $\ex{X\sim \bD}{X_J} \geq \max_j \ex{X\sim \bD}{X_j} - \alpha$ with probability at least $99/100$. This probability is taken over the randomness of the samples observed by $M$ and the algorithm $M$ itself.
\end{defn}

The following theorem can be found in \cite{Cheu21}.

\begin{thm}
If $M$ is an $(\eps,\delta)$-internally private algorithm that solves $(\alpha,d)$-selection and $\delta \log (d/\delta) \ll \alpha^2 \eps^2 / d$, then its sample complexity is $n=\Omega(\sqrt{d}/\alpha\eps)$.
\end{thm}

By combining the above with Theorem \ref{thm:main}, we arrive at the following bound on the sample complexity of any feature selector in the (non-interactive) multi-server model.

\begin{thm}
If $\Pi$ is a $k$-server protocol that solves $(\alpha,d)$-selection with sample complexity $n$ and offers $(\eps,\delta)$-differential privacy against $\lceil k/2\rceil$ corrupt servers for $\delta=0$ or $\delta \log (d/\delta) \ll \alpha^2 \eps^2 / d$, then $n=\Omega(\sqrt{d}/\alpha\eps)$.
\end{thm}

The celebrated exponential mechanism by McSherry and Talwar \cite{MT07} implies a centrally private sample complexity that is only logarithmic in $d$.

\subsection{Simple Hypothesis Testing}
\begin{defn}[$d$-Wise Simple Hypothesis Testing]
Let $d$ be any integer larger than 1 and let $\alpha$ be any real in the interval $(0,1/2)$. An algorithm $M$ solves \emph{$d$-wise simple hypothesis testing with error $\alpha$ and sample complexity $n$} if, for any set of $d$ distributions $\cP$ satisfying  $\sd{\bD}{\bD'}\geq \alpha$ for every distinct pair $\bD,\bD' \in \cP$, when given $n$ independent samples from an arbitrary $\bD \in \cP$ as input, the algorithm outputs $\bD$ with probability $\geq 99/100$. This probability is over the randomness of the samples observed by $M$ and $M$ itself.
\end{defn}

The following theorem can be found in \cite{Cheu21}.

\begin{thm}
If $M$ is an $(\eps,\delta)$-internally private algorithm that solves $d$-wise simple hypothesis testing with error $\alpha$ and either $\delta=0$ or $\delta \log (d/\delta) \ll \alpha^2 \eps^2 / d$, then its sample complexity is $n=\Omega(\sqrt{d}/\alpha\eps)$.
\end{thm}

By combining the above with Theorem \ref{thm:main}, we arrive at the following bound on the sample complexity of any $d$-wise simple hypothesis tester in the (non-interactive) multi-server model.

\begin{thm}
If $\Pi$ is a $k$-server protocol that solves $d$-wise simple hypothesis testing with error $\alpha$ and offers $(\eps,\delta)$-differential privacy against $\lceil k/2\rceil$ corrupt servers for $\delta=0$ or $\delta \log (d/\delta) \ll \alpha^2 \eps^2 / d$, then $n=\Omega(\sqrt{d}/\alpha\eps)$.
\end{thm}

Work by Bun, Kamath, Steinke, and Wu \cite{BunKSW19} contains a centrally private algorithm with a logarithmic sample complexity.

\subsection{Uniformity Testing}

\begin{defn}[Uniformity Testing]
\label{def:ut}
An algorithm $M$ solves $\alpha$-\emph{uniformity testing} with sample complexity $n$ when:
\begin{itemize}
    \item If $\vec{x}\sim \bU^n$, then $\pr{}{M(\vec{x}) = \textrm{``uniform''}} \geq 2/3$, and
    \item If $\vec{x}\sim \bD^n$ where $\sd{\bD}{\bU} > \alpha$, then $\pr{}{M(\vec{x}) = \textrm{``not uniform''}} \geq 2/3$
\end{itemize}
where the probabilities are taken over the randomness of $M$ and $\vec{x}$.
\end{defn}

The following is implied by the proof of Theorem 3 from Amin et al. \cite{AJM19}.
\begin{thm}
For $\alpha < 1/2$, any $\eps$-internally private $\alpha$-uniformity tester has sample complexity
$$n=\Omega\left(\frac{d^{2/3}}{\alpha^{4/3} \eps^{2/3}}+ \frac{\sqrt{d}}{\alpha^2}+ \frac{1}{\alpha \eps}\right).$$
\end{thm}

\begin{thm}
If $\Pi$ is a $k$-server protocol that solves $\alpha$-uniformity testing and offers $\eps$-differential privacy against $\lceil k/2\rceil$ corrupt servers, then $$n=\Omega\left(\frac{d^{2/3}}{\alpha^{4/3} \eps^{2/3}}+ \frac{\sqrt{d}}{\alpha^2}+ \frac{1}{\alpha \eps}\right).$$
\end{thm}

In contrast, Acharya, Sun, and Zhang describe a centrally private algorithm whose sample complexity scales with $\sqrt{d}$ \cite{ASZ18}.


\section{Acknowledgements}
We would like to thank Matthew Joseph for correspondence that refined our understanding of Bayesian re-sampling. We also thank Kobbi Nissim for suggestions for our sample complexity analysis.

\bibliographystyle{plain}
\bibliography{refs}

\appendix
\section{A Two-Server Protocol for Robust Count Estimates}
\label{apdx:example}

Here, we sketch an example of a protocol in our model. It performs differentially private counting (summation of $\zo$ values). Steinke \cite{Steinke20} gave a simple protocol for this problem but a malicious user can greatly skew the count estimate.\footnote{Each honest user secret-shares their value across servers, so the modulus must be at least $n$. But single malicious user can send shares that encode a $\Omega(n)$ value instead of a $\zo$ value and honest servers cannot detect this.} The protocol by Talwar \cite{Talwar22} is designed with such attacks in mind. It performs high-dimensional addition (summation of values in the unit $\ell_2$ ball). The one-dimensional nature of counting admits a greatly simpler construction.

\medskip

Before we define the algorithms that make up the protocol, we give some preliminary notation. For predicate $p$, $\ind{p}$ is 1 if $p$ is true and 0 if it is false. For natural number $t$, let $\bD_t$ be the distribution over $[t]$ such that $\pr{\eta \sim \bD_t}{\eta = v} \propto \exp(-\eps\cdot |t/2-v|)$.

On input $x_i$, the local randomizer $R$ samples $\eta_i$ from $\bD_t$ and reports
$$
(y_{i,1} \gets x_i+\eta_i, y_{i,2}\gets \eta_i)
$$
The first server samples $\alpha_1$ from $\bD_t$ and reports
$$
z_{1\to 2} := \alpha_1 + \sum_{i\in [n]} y_{i,1}\cdot \ind{y_{i,1} \in [0,t+1]}
$$
to the second server, who then reports
$$
z_\out := z_{1\to 2}+ \alpha_2 - t - \sum_{i\in [n]} y_{i,2}\cdot \ind{y_{i,2} \in [0,t+1]}
$$
where $\alpha_2$ is yet another sample from $\bD_t$.

\medskip

In our analysis, we make the simplifying assumption that $\delta \ll \eps < 1$.

\begin{clm}
For $t = \Theta(\frac{1}{\eps}\log\frac{1}{\delta})$, the protocol is $(\eps,\delta)$-differentially private against 1 (semi-honest) corrupted server.
\end{clm}
\begin{proof}[Proof Sketch]
Without loss of generality, we will ensure privacy for user 1 and assume the adversary corrupts all other users.

If server 1 is corrupted but server 2 is honest, the only variables pertaining to user 1 the adversary can obtain are $y_{1,1}$ and $\alpha_2 + y_{1,1} - y_{1,2} = \alpha_2 + x_1$. The former is received directly from user 1. The latter is obtainable by subtracting the other user's messages from $z_\out$. $y_{1,1}$ is the result of adding noise from a truncated discrete Laplace distribution to $x_1$, a value with sensitivity 1. The same is true for $\alpha_2 + x_1$. Hence we have $(\eps,\delta)$-differential privacy from composition (and the right choice of parameters).

If server 2 is corrupted (but server 1 is honest), the adversary can observe $y_{1,2}$ and the value $\alpha_1 + y_{1,1}$. The former is direct from user 1 while the latter is obtained by subtracting the other user's messages from $z_{1\to 2}$. The adversary can compute $\alpha_1 + y_{1,1} - y_{1,2} = \alpha_1 + x_1$ which is $(\eps,\delta)$-differentially private for the same reason that $\alpha_2+x_1$ is private.
\end{proof}

\begin{clm}
If all parties are honest and $t$ is set as above, then the protocol produces an unbiased estimate of the count such that, with 90\% probability, the error is $O\paren{\frac{1}{\eps}}$. If there are $m$ malicious users and no malicious servers, the error is $O\paren{\frac{m}{\eps}\log \frac{1}{\delta}}$.
\end{clm}
\begin{proof}[Proof Sketch]
We first argue that there is no bias in the honest execution. In this case, we equate $z_\out$ with the sum of $h_1 = \alpha_1+\alpha_2-t$ and $h_2 = \sum_{i\in [n]} y_{i,1}\cdot \ind{y_{i,1} \in [0,t+1]} - y_{i,2}\cdot \ind{y_{i,2} \in [0,t+1]}$. The random variable $h_1$ has mean 0 because $t=\ex{}{\alpha_1+\alpha_2}$. And observe that the construction of $y_{i,1},y_{i,2}$ implies $h_2 = \sum x_i$.

Now we argue the error is likely low. By manipulating geometric series, it can be shown that $|\alpha_1-t/2|$ is at most $k$ with probability $(e^\eps+1-e^{-\eps k})/(e^\eps+1-\Theta(\delta))$. Invoking our bounds on $\delta$ and $\eps$, this probability is at least 0.95 when $k=\Theta(1/\eps)$. The same goes for $\alpha_2$. A union bound completes the proof.

We conclude with the analysis of the manipulation case. The honest servers limit the influence of any user on the output to be $O(t)$ because they only add messages that belong in the range $[0,t+1]$. Hence, no coalition of $m$ users can introduce more than $O\paren{\frac{m}{\eps}\log \frac{1}{\delta}}$ bias.
\end{proof}


\section{Deferred Proofs}
\label{apdx:deferred}

\begin{lem}[Copy of \ref{lem:geo-chernoff}]
Fix any $p_1,\dots,p_n, \ell \in (0,1/2]$ such that $\ell \leq p_i$ for every $i\in[n]$. If we sample $\eta_i$ from $\Geo(p_i)$ for every $i\in [n]$, then
$$
\sum_{i=1}^n \eta_i = O\paren{ \frac{n}{\ell}\ln \frac{1}{\ell} + \frac{1}{\ell}\ln \frac{1}{\beta} }
$$
with probability at least $1-\beta$, for any $\beta \in (0,1)$.
\end{lem}
\begin{proof}
We use a Chernoff-style technique. For any $t,v>0$,
\begin{align*}
\pr{\eta_i \sim \Geo(p_i)}{\sum_{i=1}^n \eta_i > v} &\leq \exp(-vt)\cdot \ex{\eta_i \sim \Geo(p_i)}{\exp\paren{t\sum_{i=1}^n \eta_i}} \\
    &= \exp(-vt)\cdot \prod_{i=1}^n \ex{\eta_i \sim \Geo(p_i)}{\exp(t\cdot \eta_i)} \stepcounter{equation} \tag{\theequation} \label{eq:geo-chernoff-1}
\end{align*}
The upper bound comes from Markov's inequality and the equality comes from independence.

Suppose we set $t=\ln \frac{1-\ell/2}{1-\ell}$. Note that it lies in the interval $(0, \ln \frac{1}{1-\ell})$. In turn, observe $(0, \ln \frac{1}{1-\ell})$ is a subset of the interval $(0, \ln \frac{1}{1-p_i})$ for any $i$. This means the MGF of $\Geo(p_i)$ is well-defined for our $t$:
\begin{align*}
\ex{\eta_i \sim \Geo(p_i)}{\exp(t\cdot \eta_i)} &= p_i e^t\cdot \frac{1}{1-(1-p_i)e^t}\\
    &= p_i e^t\cdot \frac{1}{1-(1-p_i)\frac{1-\ell/2}{1-\ell}} \\
    &\leq p_i e^t\cdot \frac{2}{\ell} \tag{$p_i\geq \ell$} \\
    &= p_i \cdot \frac{1-\ell/2}{1-\ell} \cdot \frac{2}{\ell}\\
    &\leq \frac{1-\ell/2}{1-\ell} \cdot \frac{1}{\ell} \tag{$p_i\leq 1/2$}
\end{align*}
Hence,
\begin{align*}
    \eqref{eq:geo-chernoff-1} &\leq \exp(-vt)\cdot \paren{ \frac{1-\ell/2}{1-\ell}  }^n \cdot \paren{\frac{1}{\ell}}^n\\
    &= \paren{ \frac{1-\ell/2}{1-\ell}  }^{n-v} \cdot \paren{\frac{1}{\ell}}^n  \stepcounter{equation} \tag{\theequation} \label{eq:geo-chernoff-2}
\end{align*}

Let $b= \frac{1-\ell/2}{1-\ell}$, and $v= n+ n\log_b (1/\ell) + \log_b (1/\beta)$. By substitution,
\begin{align*}
\eqref{eq:geo-chernoff-2} &= b^{n-v} \cdot \paren{\frac{1}{\ell}}^n\\
    &= \beta
\end{align*}

We now bound $v$ by an easier-to-read expression:
\begin{align*}
v &= n+ n\log_b \frac{1}{\ell} + \log_b \frac{1}{\beta}\\
    &=n + \frac{n\ln (1/\ell) + \ln(1/\beta)}{\ln ((1-\ell/2)/(1-\ell))} \tag{Change of base}\\
    &\leq n + \frac{n\ln (1/\ell) + \ln(1/\beta)}{\ell/2}\\
    &= n + \frac{2n}{\ell}\ln \frac{1}{\ell} + \frac{2}{\ell}\ln \frac{1}{\beta}
\end{align*}
This concludes the proof.
\end{proof}

\end{document}